\documentclass[aps, prx,reprint, superscriptaddress]{revtex4-1}
\usepackage{graphicx}
\usepackage{amsmath}
\usepackage{color}
\usepackage{dsfont, bm,  units}
\usepackage{dcolumn}

\newcommand{\partialdif}[2]{\frac{\partial #1}{\partial #2}}
\newcommand{\Te}[0]{T_{\mathrm{e}}}
\newcommand{\Tp}[0]{T_{\mathrm{p}}}

\usepackage{soulutf8}
\usepackage{upgreek}

\newcommand{\noneqmodel}[0]{microscopic out-of-equilibrium dynamics model}

\begin{document}


\title{Theory of out-of-equilibrium electron and phonon dynamics in metals after ultrafast laser excitation}
\author{Ulrike Ritzmann}
\affiliation{Department of Physics and Astronomy, Uppsala University, Box 516,  S-75120 Uppsala, Sweden}
\affiliation{Dahlem Center of Complex Quantum Systems, Freie Universit\"at Berlin, Arnimallee 14,  D-14195 Berlin, Germany}
\affiliation{Department of Physics, Freie Universit\"at Berlin, Arnimallee 14,  D-14195 Berlin, Germany}
\author{Peter M. Oppeneer}
\affiliation{Department of Physics and Astronomy, Uppsala University, Box 516, S-75120 Uppsala, Sweden}
\affiliation{Department of Physics, Freie Universit\"at Berlin, Arnimallee 14,  D-14195 Berlin, Germany}
\author{Pablo Maldonado}
\affiliation{Department of Physics and Astronomy, Uppsala University, Box 516, S-75120 Uppsala,  Sweden}
\date{\today}

\begin{abstract}
The out-of-equilibrium dynamics of electrons and phonons upon laser excitation are often described by the two-temperature model, which assumes that both subsystems are separately in thermal equilibrium. However, recent experiments show that this description is not sufficient to describe the out-of-equilibrium dynamics on ultrashort timescales. Here, we extend and apply a parameter-free microscopic out-of-equilibrium model to describe the ultrafast laser-induced system dynamics of archetypical
metallic systems such as gold, aluminum, iron, nickel, and cobalt.
We report strong deviations from the two-temperature model on the picosecond timescale for all the materials studied, even for those where the assumption of separate thermal equilibriums seemed less restrictive, like in gold. Furthermore, we demonstrate the importance of the phonon-mode dependent electron-phonon coupling for the relaxation process and reveal the significance of this channel in the lattice equilibration through an indirect coupling between phonons via the electronic system.

\end{abstract}

\pacs{} 

\maketitle

\section{Introduction}
The development of new experimental techniques using ultrashort laser pulses has allowed in recent years to gain access to novel phenomena taking place at subpicosecond timescales when the system is still strongly out-of-equilibrium  \cite{Sun1993, Groeneveld1995, Stojchevska2014, Rameau2016} with heavily intertwined degrees of freedom. 
Thus, the complex interplay of electronic, phononic and spin degrees of freedom after laser excitation has become the focus of contemporary research which has led to the discovery of new phenomena such as ultrafast demagnetization \cite{Beaurepaire1996, Bigot2009, Kirilyuk2010}, change of magnetic anisotropy \cite{Hansteen2005}, ultrafast generation of lattice strain waves \cite{Kim2012, Henighan2016}, coherent phonon generation \cite{Harmand2013, Lindenberg2017}, laser-induced superconductivity  at high temperatures \cite{Mitrano2016}, excitation of ultrafast spin currents \cite{Rudolf2012} and of high-frequency exchange magnons \cite{Razdolski2017}, spintronic THz emitters \cite{Seifert2016} or femtosecond activation of magnetoelectricity \cite{Bossini2018}.

From a theoretical perspective, thermal models such as the two- and three-temperature model are commonly used to describe the ultrafast interplay of electrons, phonons and spins. These models are based on the assumption that the laser excites a thermalized electron distribution and that electrons and phonons are each individually in thermal equilibrium during the relaxation, and were originally derived by Kaganov \textit{et al.}\ \cite{Kaganov1957} and later extended by Anisimov \textit{et al.}\ \cite{Anisimov1974} for the case of phonons and electrons subsystems. The assumption implies that the electron system thermalizes instantaneously through electron-electron interaction and similarly, the phonon system thermalizes immediately through phonon-phonon interaction or that the electron-phonon coupling and specific heat is independent on the phonon modes. The time evolution of the two subsystems can then be described by the time dependence of the electron temperature $\Te$ and the phonon temperature $\Tp$ derived from the following coupled equations:
\begin{align}
\label{eq_2TM1}
C_\mathrm{e}\partialdif{\Te^\mathrm{2TM}}{t}=&-G_\mathrm{ep}\left(\Tp^\mathrm{2TM}-\Te^\mathrm{2TM}\right)+P(t)\mbox{,}\\
\label{eq_2TM2}
C_\mathrm{p}\partialdif{\Tp^\mathrm{2TM}}{t}=&G_\mathrm{ep}\left(\Tp^\mathrm{2TM}-\Te^\mathrm{2TM}\right)\mbox{.}
\end{align}
The specific heat constants for electrons, $C_\mathrm{e}$, and phonons, $C_\mathrm{p}$, as well as the electron-phonon coupling constant $G_\mathrm{ep}$ can be determined either experimentally or by first-principles calculations  \cite{Allen1987, Lin2008}. $P(t)$ represents the absorbed power from the laser excitation.

However, recent experimental and theoretical works have evidenced that ultrashort laser pulses induce strong out-of-equilibrium dynamics that cannot be described by simplified thermal models in the subpicosecond time regime \cite{Henighan2016, Waldecker2016,Waldecker2017}. 
As a consequence, a large number of theoretical and experimental studies have tried to shine light on the laser-induced out-of-equilibrium dynamics \cite{Aoki2014, Baranov2014, Carpene2006, Rethfeld2002, Bauer2015,  Murakami2015, Ono2017, Abdurazakov2018, Klett2018, Tamm2018, Reid2018}. 
New theoretical models have been derived to improve the modelling on ultrashort timescales. One approach is to consider distinct phonon-branch dependent dynamics \cite{Waldecker2016}. Further approaches introduce mode-dependent couplings and consider out-of-equilibrium distributions for electrons or phonons \cite{Maldonado2017, Weber2018a, Sadasivam2017}. 

Here, we use a \noneqmodel\ derived by Maldonado \textit{et al.}\ to investigate laser-induced dynamics of the phononic system in different metallic materials on picosecond timescales based on first-principles calculations \cite{Maldonado2017}. The model includes phonon-mode dependent electron-phonon and phonon-phonon coupling. We discuss the out-of-equilibrium dynamics in gold, aluminum, nickel, iron, and cobalt. 
Our results reveal strong deviations from those provided by the two-temperature model even a long time after the laser excitation, evidencing thus the failure of the two-temperature model at these timescales. We demonstrate the relevance of the mode-dependent electron-phonon coupling and show the distinct dynamical behavior of the different materials studied. Furthermore, we show that electron-phonon coupling has a significant role in the lattice equilibration mechanism, specifically, as a channel to transfer heat from hot to cold phonon modes via the electronic system.

\section{\noneqmodel}
In the following we will describe the theoretical model which provides the out-of-equilibrium system dynamics triggered by an ultrashort laser pulse in different metals, showing additionally the microscopic mechanisms involved in the equilibration process. 
Here it is important to mention that although the out-of-equilibrium electron dynamics is of special relevance to understand different laser-induced ultrafast phenomena, such as ultrafast demagnetization \cite{Beaurepaire1996} or energy relaxation in strongly correlated electronic systems \cite{Konstantinova2018}, it has been shown recently that the electronic relaxation time is only about a few tens of femtoseconds \cite{Gierz2015,Rohde2018,Kemper2018,Tengdin2018}. Considering that the relevant timescales of this work span several tens of picoseconds, we can safely assume a thermal distribution of the electronic system during the whole process with a transient electron temperature $T_{\rm e}(t)$. Contrarily, the lattice dynamics needs an explicit out-of-equilibrium representation of the phonon population \cite{Maldonado2017}. To achieve this aim, we use the ansatz that the out-of-equilibrium phononic distribution function can be written as
%
\begin{align}
n_Q(t)&=\frac{1}{\exp\left(\frac{\hbar\omega_Q}{k_\mathrm{B}T(t)}+\Psi_Q(t)\right)-1} ,
\end{align}
where $\omega_Q$ is the frequency of the phonon mode, $Q\equiv(\mathbf{q},\nu)$, with $\mathbf{q}$ being the wavelength and $\nu$ the phonon branch. $T(t)$ is the lattice temperature and $\Psi_Q$ a measurement of the deviation from equilibrium \cite{Srivastava1990,Maldonado2018}. Correspondingly, we can rewrite the above expression with a more compact notation by defining a phonon mode-dependent effective phonon temperature, $\Tp^{Q}(t)$, that would differ for distinct regions of the BZ,
\begin{align}
 n_Q(t)&=\frac{1}{\exp\left(\frac{\hbar\omega_Q}{k_\mathrm{B}\Tp^{Q}(t)}\right)-1} .
\end{align}

These branch and wavevector-dependent phonon temperatures have their counterparts for the electronic states at different regions of the Fermi surface as observed experimentally by Schutt \textit{et al.}\ \cite{Schutt2018}. Then, by using the conservation of total energy and the Boltzmann kinetic theory, a set of coupled equations of motion for the temperatures of each phonon mode $Q$ and of the electrons can be derived:
\begin{align}
\label{eq_dyn1}
C_\mathrm{e}\partialdif{\Te}{t}=&\sum_{Q}{\gamma_QC_QI(\Tp^Q-\Te)[1+J(\Tp^Q-\Te)]}+P(t),\\
\label{eq_dyn2}
C_{Q}\partialdif{\Tp^Q}{t}=&-\gamma_QC_QI(\Tp^Q-\Te)[1+J(\Tp^Q-\Te)]\nonumber\\&-\sum_{k'}{C_Q\Gamma_{Qk'}(\Tp^Q-\tilde{\Tp}^{k,k'})} \times \nonumber\\&~~~~[1+J(\Tp^Q-\tilde{\Tp}^{k,k'})] ,
\end{align}
where $C_\mathrm{e}$ and $C_{Q}$ are the electronic and phonon-mode $Q$ heat capacities, and $\gamma_Q$ and $\Gamma_{Qk'}$ are the phonon linewidths due to electron-phonon and phonon-phonon scattering, respectively.
These parameters are obtained from first principles, in our case on a dense grid of $20^3$ $k$-points.
Details of the derivation of these equations can be found in the work of Maldonado \textit{et al.}\ \cite{Maldonado2017}. However, it is important to mention that the model has been further improved in this work to explicitly avoid the use of the single relaxation approximation in the derivation of Equations \eqref{eq_dyn1} and \eqref{eq_dyn2} (further details can be found in the Appendix).  

The laser excitation is considered to have a gaussian distribution with the rate of the absorbed energy given by
		\begin{align}
		P(t)=(1-R)\, \phi \, \alpha\frac{2}{s\sqrt{2\pi}}\exp{\left(-\frac{-2t^2}{s^2}\right)} .
		\end{align}
The full width at half maximum of the laser pulse $\sigma$ is 100 fs and given by $\sigma=\sqrt{s/(2\cdot \sqrt{2\cdot \ln(2)}\,)}$.  We consider a laser fluence of $\phi=50$ mJ/cm$^2$. $R$ is the reflectivity and $\alpha$ is the adsorption coefficient. For a better comparison of the results obtained for the different metallic systems here studied, we use the same rate of absorbed energy for all cases. The amount of absorbed energy is defined initially for gold, and then taken as reference for the other systems. In this case, the reflectivity is given by $R=0.98$ and the adsorption coefficient is $\alpha=0.80916$\,cm$^{-1}$ for a wavelength of the laser pulse of 800 nm. Note that the adsorption coefficients can vary largely for the materials and therefore, the laser fluences have been modified accordingly.

\section{Results}
Before analyzing the out-of-equilibrium dynamics obtained as a solution of Equations \eqref{eq_dyn1} and \eqref{eq_dyn2} for the different metallic systems here studied, i.e.\ gold, aluminum, iron, nickel, and cobalt, we have computed the phonon-mode dependent linewidths using the ABINIT software \cite{Gonze2009} to extract the electron-phonon couplings and the system's force constants. We have also calculated the electron-phonon coupling constant $G_\mathrm{ep}=\sum_Q{\gamma_QC_Q}$, commonly used as a free parameter in the two-temperature model. The \textit{ab initio} results for the different materials are summarized in Table \ref{table_parameter} computed at $T = 300$\,K, along with averaged values for the phonon linewidth due to electron-phonon coupling, $\langle \gamma_Q \rangle=1/N_Q\sum_Q{\gamma_Q}$, and the averaged value for the phonon linewidth due to phonon-phonon coupling, $\langle \Gamma_{Qk'} \rangle=1/N_{k'}\sum_{k'}{\Gamma_{Qk'}}$.


Importantly, we obtain that in gold and aluminum the role of anharmonicities is more significant, indicated by the larger phonon-phonon linewidth and phonon-phonon coupling as compared to nickel, iron, and cobalt.  Moreover, in the case of gold this coupling is even larger than the electron-phonon coupling strength, and therefore  a fast lattice equilibration could be expected, which would justify the use of the two-temperature model. Contrarily, we find for the rest of the systems studied that the electron-phonon strength coupling is larger than the phonon-phonon coupling strength, and therefore it could be already expected that the lattice cannot be modeled with a thermal distribution having a single lattice temperature. The values of the electron-phonon coupling parameters reported in Table \ref{table_parameter} are in good agreement with previously published works, except for the case of iron where a large deviation is observed  \footnote{We compared our values with http://www.faculty.virginia.edu/CompMat/electron-phonon-coupling}. In contrast to our work, where we use a first-principles computation of the parameters, the electron-phonon coupling parameter  was previously estimated by using a Debye approximation, to which we associate the discrepancies with our results.

\begin{table}[t]
\begin{ruledtabular}
\caption{\label{table_parameter} \textit{Ab initio} calculated material specific coupling constants. Given are the averaged linewidth for electron-phonon coupling, $\langle\gamma_Q\rangle$, and for phonon-phonon coupling, $\langle\Gamma_{Qk'}\rangle$, and the corresponding electron-phonon coupling constant $G_\mathrm{ep}$.}
\begin{tabular}{c|c|c|c}
Material & $\langle\gamma_Q\rangle$  (GHz)& $\langle\Gamma_{Qk'}\rangle$ (GHz) & $G_\mathrm{ep}$ ($10^{17}$W/(m$^3$K))\\[0.1cm]
\hline
Gold & 4.7  & 90.6  & 0.225\\
Aluminum & 101.1 & 77.2 &  4.47\\
Iron &   164.0 & 6.7 & 10.5\\
Nickel & 272.5 &  27.0 & 18.9\\
Cobalt & 458.7 & 16.8 & 33.4\\
\end{tabular}
\end{ruledtabular}
\end{table}

Here, it is important to stress that our model has been derived under the assumption that the spin subsystem plays a negligible role in the lattice dynamics due to the usually small magnetic heat capacities that lead to a small energy adsorption in the spin systems. Although this assumption is rigorously valid for nonmagnetic materials such as Au and Al it is only valid at low temperatures for magnetic systems and it might fail close to the Curie temperature. On the other hand, we include an explicit dependence of the electron-phonon coupling on the electron spins (the total electron-phonon coupling is the result of the sum of the coupling for minority and majority electrons). Here, we consider the spin-dependent values for fully magnetized systems and do not include magnetization dependent spin densities. Although we assume that the inclusion of the spin subsystem in our model will modify the quantitative behavior of the dynamics, the qualitative behavior would be the same, and therefore the timescales of the dynamics and conclusions drawn in this work remain valid.

\subsection{Out-of-equilibrium dynamics in gold}
\begin{figure}[t]
	\centering
	\includegraphics[width=0.48\textwidth]{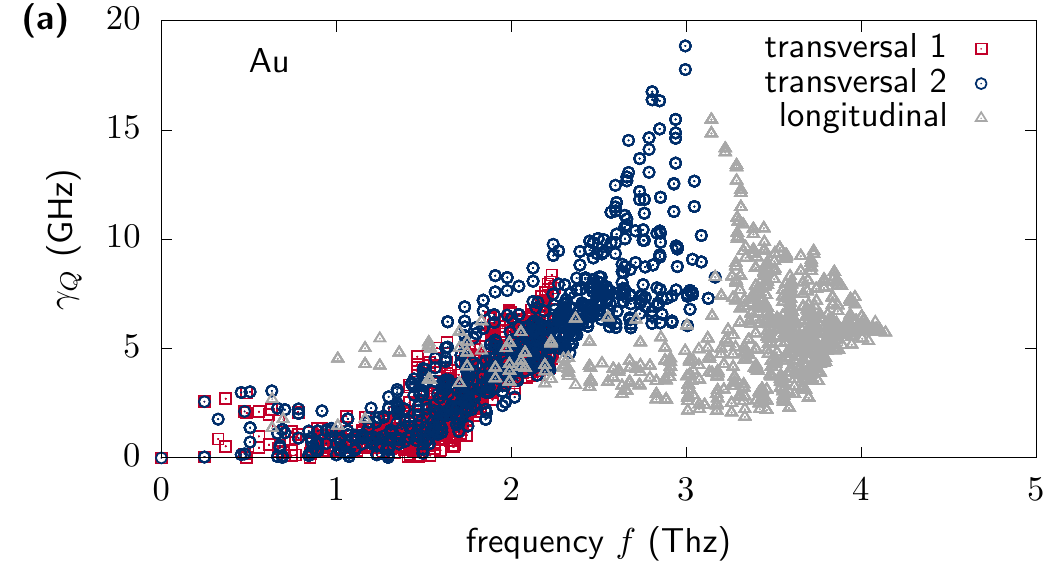}
	\includegraphics[width=0.48\textwidth]{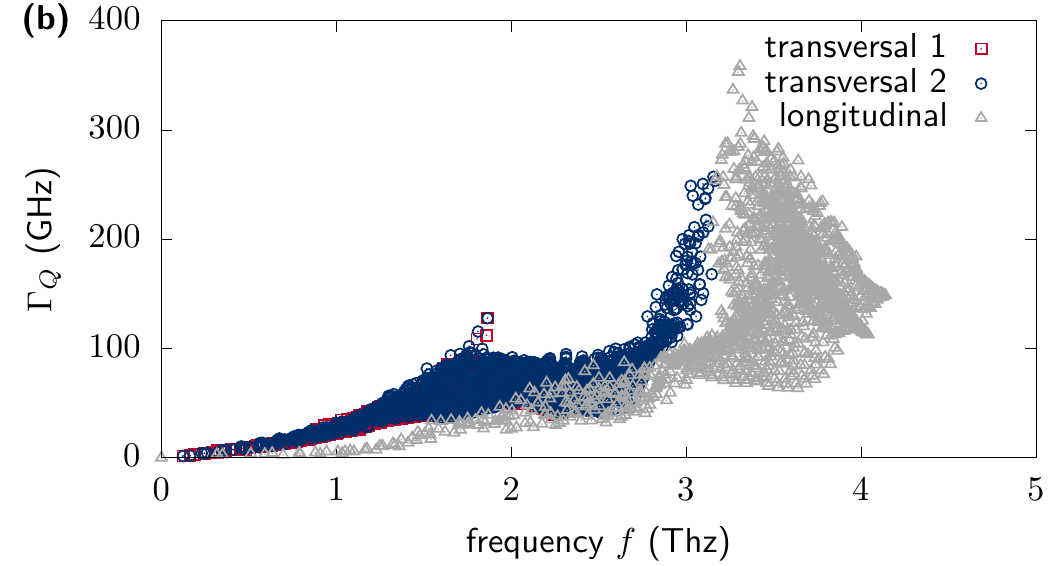}
	\caption{\label{fig_gold_coupling} \textit{Ab initio} calculated electron-phonon linewidth (top) and phonon-phonon linewidth (bottom) in gold as a function of the phonon frequency for the three phonon branches.}
\end{figure}%
One of the main assumptions of the two-temperature model is that phonons are in complete thermal equilibrium during the interaction with the electrons. 
This condition can be fulfilled either by assuming an homogeneous electron-phonon coupling such that the heat is homogeneously distributed or by assuming a phonon-phonon interaction strength much larger than the electron-phonon coupling strength such that the lattice is equilibrated instantaneously. 
As for the case of Au, the former condition is not fulfilled as we can see in Figure \ref{fig_gold_coupling}(a) where we show the \textit{ab initio} calculated mode-dependent phonon linewidth due to electron-phonon interaction. The broad dispersion of linewidths along the phonon frequencies evidences how the energy will be largely inhomogeneously distributed among the different phonon modes. On the other hand, the phonon linewidths due to the phonon-phonon interaction, which are  also strongly mode dependent, present values that are one order of magnitude larger than those stemming from the electron-phonon scatterings, as illustrated in Figure \ref{fig_gold_coupling}(b). This potentially suggest the validity of a thermal description of the lattice in the two-temperature model for Au. Quantitatively this is also suggested by the averaged electron-phonon linewidth which has a value of about 4.7\,GHz, corresponding to an averaged lifetime of 106\,ps, whereas the averaged phonon-phonon linewidth is 90.6\,GHz corresponding to a lifetime of 5.5\,ps.


\begin{figure}[t]
	\centering
	\includegraphics[width=0.48\textwidth]{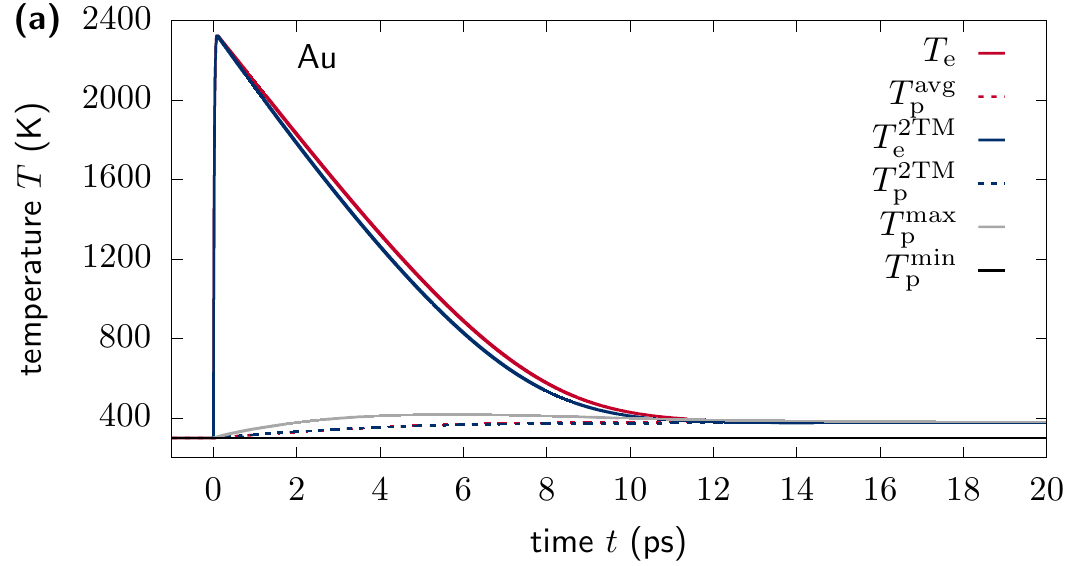}
	\includegraphics[width=0.48\textwidth]{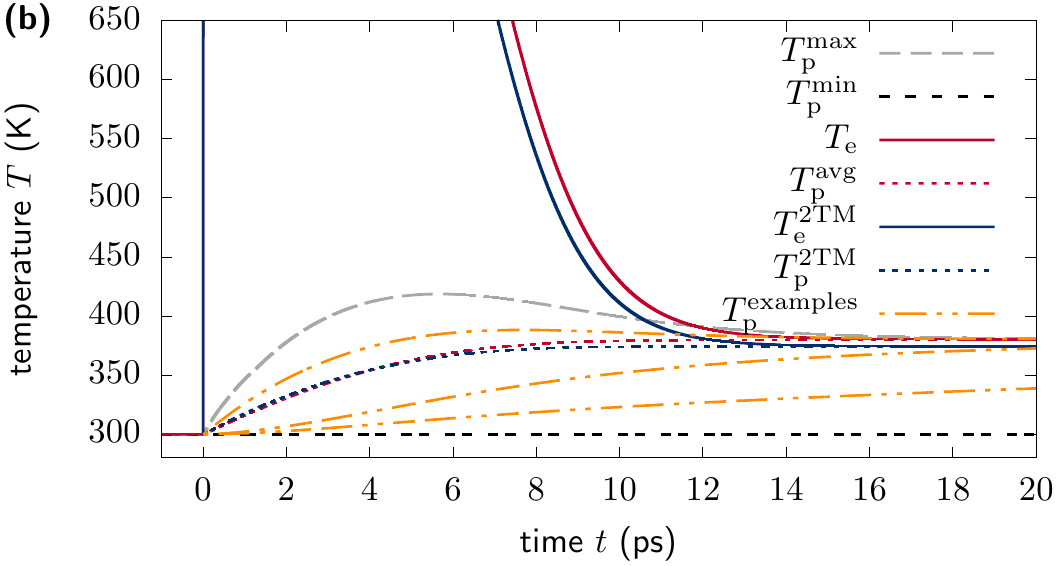}
	\caption{\label{fig_gold_2TM}Electron and phonon temperatures  computed as a function of time in gold. (a) Comparison of the  temperatures  computed from the two-temperature model (labeled 2TM) with those of the \noneqmodel\ as a function of time after laser excitation. (b) Time dependence of exemplary phonon modes illustrating different relaxation dynamics. 
	}
\end{figure}

We compute the dynamics of the electron and phonon systems, that are initially at 300\,K, after laser excitation by using the two-temperature model, given by equations (\ref{eq_2TM1}) and (\ref{eq_2TM2}), as well as by the \noneqmodel\ described by the equations (\ref{eq_dyn1}) and (\ref{eq_dyn2}).  To compare the two models, we calculate the averaged phonon temperature, as well as the minimum and maximum effective temperature of the phonons. The results are shown in Figure \ref{fig_gold_2TM}(a). The dynamics of the electron temperature and the averaged effective phonon temperature of the \noneqmodel\ are very similar to the predictions from the two-temperature model when using our computed \textit{ab initio} parameters. The electron subsystem is heated up to almost 2400\,K within the first picosecond and afterwards the system equilibrates within about 12\,ps having a final temperature of about 380\,K.  The electron and lattice temperature evolution in the \noneqmodel\ are slightly faster than in the two-temperature model. 

Furthermore, the large inhomogeneous energy distribution due to the strongly mode-dependent electron-phonon coupling is reflected on the very different dynamics of the maximum, minimum, and averaged effective phonon  temperatures. On  the one hand, the maximum effective temperature reaches a value of about 415\,K at $\approx$ 5 ps while the averaged effective  phonon  temperature only slowly converges to 380\,K at around 10 ps. On the other hand, a few modes remain with temperatures around 300\,K after 20\,ps, since they couple very weakly or don't couple with electrons nor with other phonons, but only to impurities which are not considered in our model. In Figure \ref{fig_gold_2TM}(b), we show the temporal evolution of effective temperatures of exemplary phonon modes with strong and weak electron-phonon coupling. Furthermore, we show that the maximum effective  phonon temperature can even exceed the electron temperature, which is a clear evidence of the lattice being  nonthermalized.   


\begin{figure}[t]
	\centering
	\includegraphics[width=0.48\textwidth]{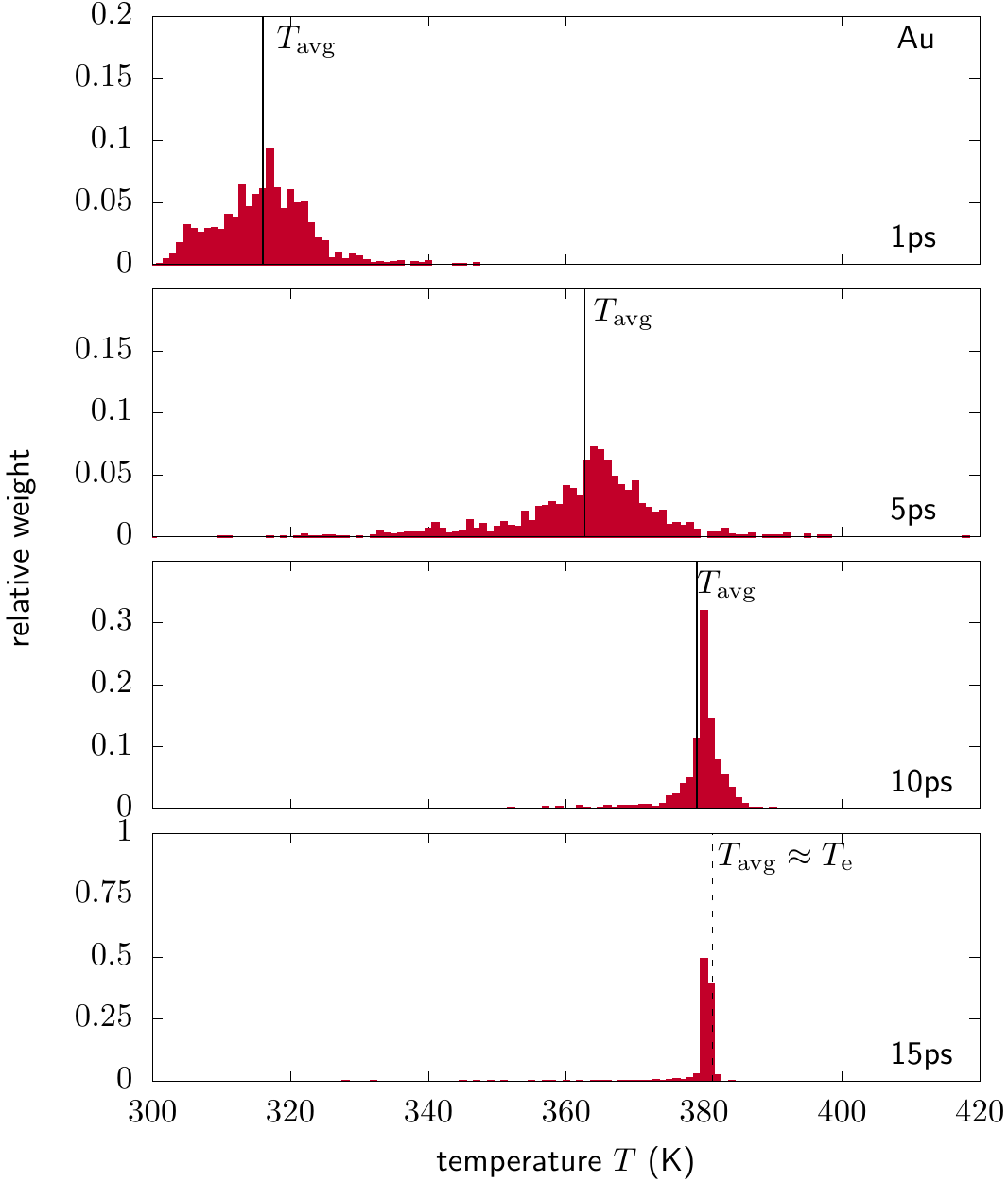}
	\caption{\label{fig_gold_hist}Phononic temperature distribution at different times during the electron-phonon relaxation in gold. The red bars show the computed relative weight of phonon frequencies having a temperature within a range of 1\,K for several different times. Vertical black lines give the average phonon temperature $T_{\rm avg}$ and the vertical dashed blue line gives the electron temperature $T_{\rm e}$. 
	}
	\end{figure}%
We gain further details of the out-of-equilibrium system dynamics by plotting in Figure \ref{fig_gold_hist} 
the relative weight of all the phonon modes at a specific temperature  within an interval of 1\,K.  At 1\,ps, we observe that the temperature range spanned by all the modes is already of about 45\,K. Since the electron temperature is still above 2000\,K at 1\,ps, phonons are further heated and their distribution broadens as shown at 5\,ps. At this time, the majority of the phonons are distributed nearly symmetrically around the averaged effective temperature within a range of 90\,K. At 10\,ps, the averaged effective temperature of the phonons is nearly converged and the electron temperature is reduced below 500\,K. 
The phonon distribution starts narrowing due to energy redistribution within the phonon system. The hottest phonons are cooling down by transferring their energy to the rest of the phonon system via phonon-phonon scatterings. At 15\,ps, the temperature distribution becomes delta-like and the phonons can be considered nearly in thermal equilibrium. Nonetheless, a few modes are still below the averaged temperature, due to their low electron-phonon and phonon-phonon coupling. This demonstrates that although phonon-phonon thermalization is very relevant  in Au, deviations from the two-temperature model still occur and a correct description of the lattice dynamics requires an out-of-equilibrium description. 

\subsection{Out-of-equilibrium dynamics in aluminum}
\begin{figure}[t]
	\centering
	\includegraphics[width=0.48\textwidth]{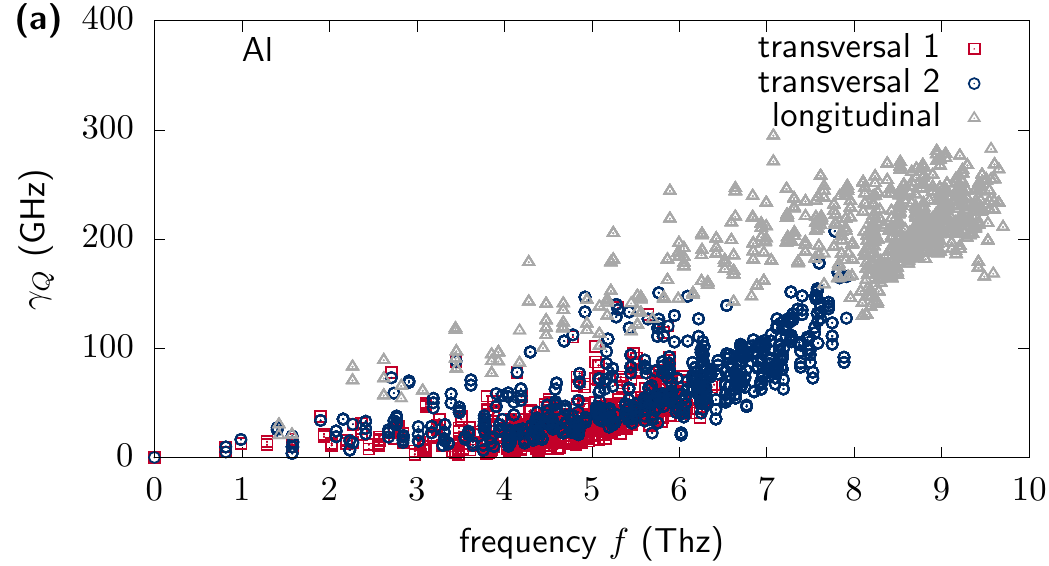}
	\includegraphics[width=0.48\textwidth]{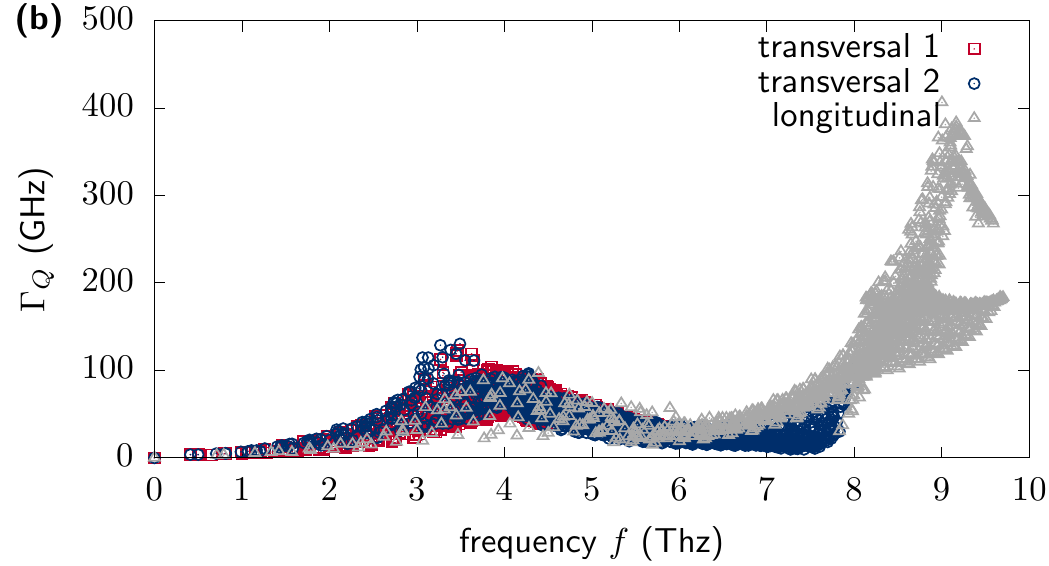}\\
	\includegraphics[width=0.48\textwidth]{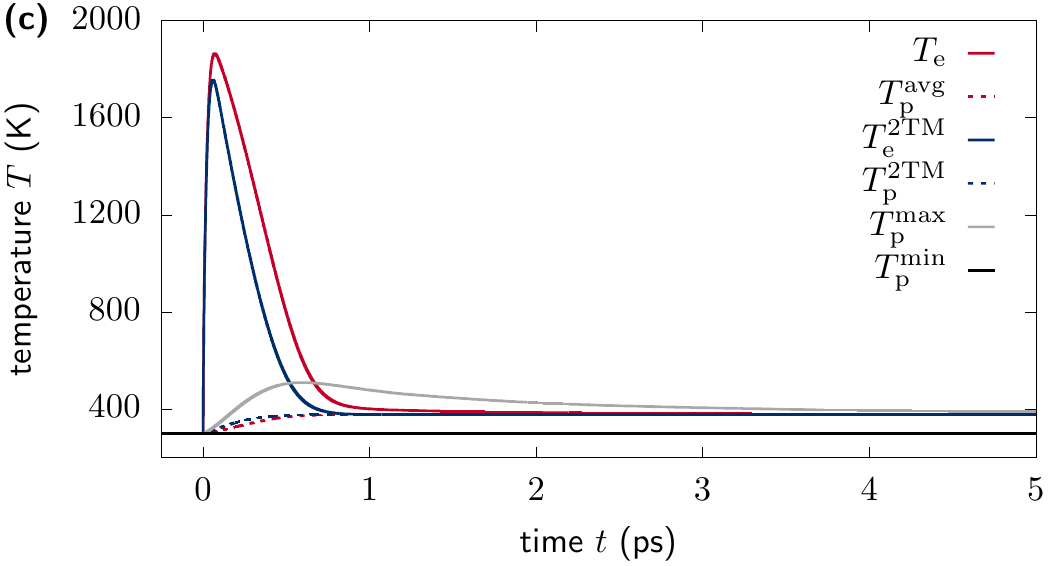}
 	\caption{\label{fig_al_2TM}(a) \textit{Ab initio} calculated electron-phonon linewidth dependent on the phonon frequency for the three phonon branches in aluminum. (b) \textit{Ab initio} calculated phonon-phonon linewidth dependent on the phonon  frequency for the three phonon branches in aluminum. (c) Comparison of the computed temperatures obtained from the two-temperature model with those of the \noneqmodel\ as a function of time after laser excitation.}
		\end{figure}%
Unlike the case of gold, and as illustrated in Table \ref{table_parameter}, aluminum has a larger electron-phonon coupling strength than the one provide by the phonon-phonon interaction, with  an averaged value of about 101\,GHz and a lifetime of 4.9\,ps. The phonon-phonon coupling is lower than in gold with an averaged linewidth of 77\,GHz corresponding to a lifetime of 6.5\,ps. The mode-dependent linewidths for both cases are shown in Figure \ref{fig_al_2TM}(a) and (b). In contrast to the dependence found in other materials, the electron-phonon coupling varies strongly for the different phonon branches. Longitudinal phonon modes couple significantly stronger to electrons and also have on average a stronger phonon-phonon coupling than the other two branches. Similar results have been shown by Tang \textit{et al.}\ \cite{Tang2010}.  
		
	\begin{figure}[t]
	\centering
	\includegraphics[width=0.48\textwidth]{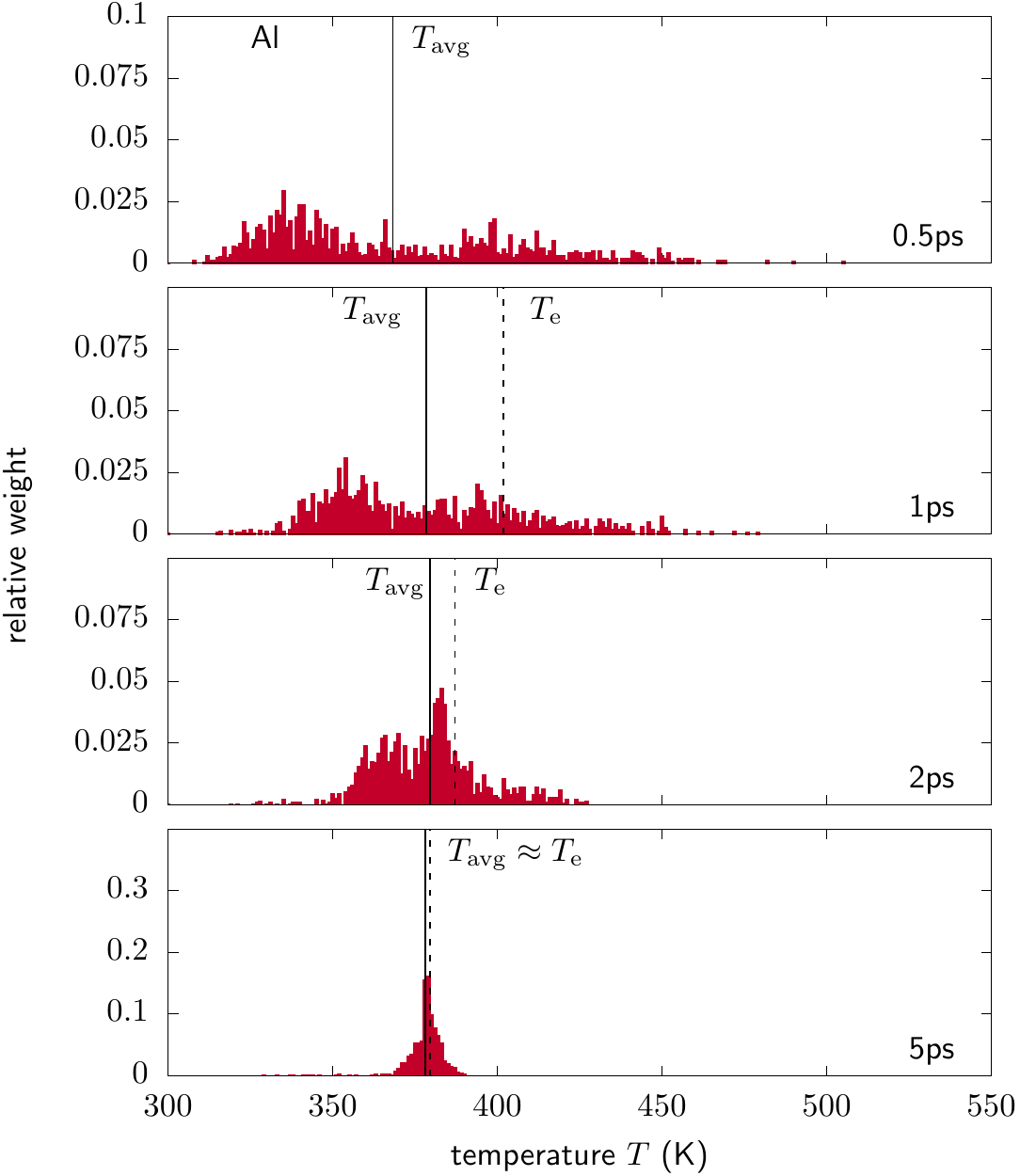}
	\caption{\label{fig_al_hist}Phononic temperature distribution at different times during electron-phonon relaxation in aluminum. The relative weight of phonon frequencies with a temperature within an interval of 1\,K is shown for different times by the red bars.}
\end{figure}

The time evolution of electrons and averaged effective phonon temperatures in comparison with predictions from the two-temperature model are shown in Figure \ref{fig_al_2TM}(c). Electrons are heated up above 1800\,K, whereas the relaxed temperature is again about 380\,K. The larger electron-phonon coupling as compared with the gold parameters, leads to a faster relaxation than in the case of gold, having the averaged effective phonon temperature and electron temperature practically converged to a common value at 1\,ps. However, it is important to emphasize that in this case we additionally observe larger deviations between the \noneqmodel\ and the two-temperature model. Thus, the electron temperature is lower in the two-temperature model compared to the \noneqmodel\ and the subsequent dynamics exhibits 
a faster relaxation. 

We also calculate the temperature distributions during the different stages of the relaxation process and the results are shown in Figure \ref{fig_al_hist}.  The phonon distribution is much broader than for gold. Additionally, some phonon modes reach an effective temperature of about 100\,K more than the averaged value. Below 2\,ps, the distribution of the phonon temperatures is still very broad and asymmetric, evidencing a strong non-equilibrium state. The broad distribution results from the independent dynamics of the different phonon branches. The longitudinal phonon modes couple stronger to the electrons and among each other, whereas the other two branches have a lower electron-phonon coupling and a lower phonon-phonon coupling. This separates their dynamics and causes a very distinct dynamics for the different branches.

At 2\,ps, the phonon distribution has narrowed. Since the electron temperature is in the range of the effective phonon temperatures, an indirect energy redistribution via electron-phonon coupling is possible. Although the electron temperature and the averaged effective phonon temperature converges at 1\,ps, the phonons can only be considered to be in thermal equilibrium after around 5\,ps. These results are in agreement with recent experimental observations and numerical results by Waldecker \textit{et al.}\ \cite{Waldecker2016}. In their work, they assign the deviations from the two-temperature model to a phonon branch-dependent coupling to the electrons, with a phenomenological phonon-phonon coupling that is fitted to reproduce the experimental equilibration. However, it is relevant to stress that our results reveal an additional strong phonon frequency-dependent behavior, which leads to an even more complex dynamics and very broad effective phonon-temperature distributions.

\subsection{Out-of-equilibrium dynamics in nickel}%
\begin{figure}[t]
	\centering
	\includegraphics[width=0.48\textwidth]{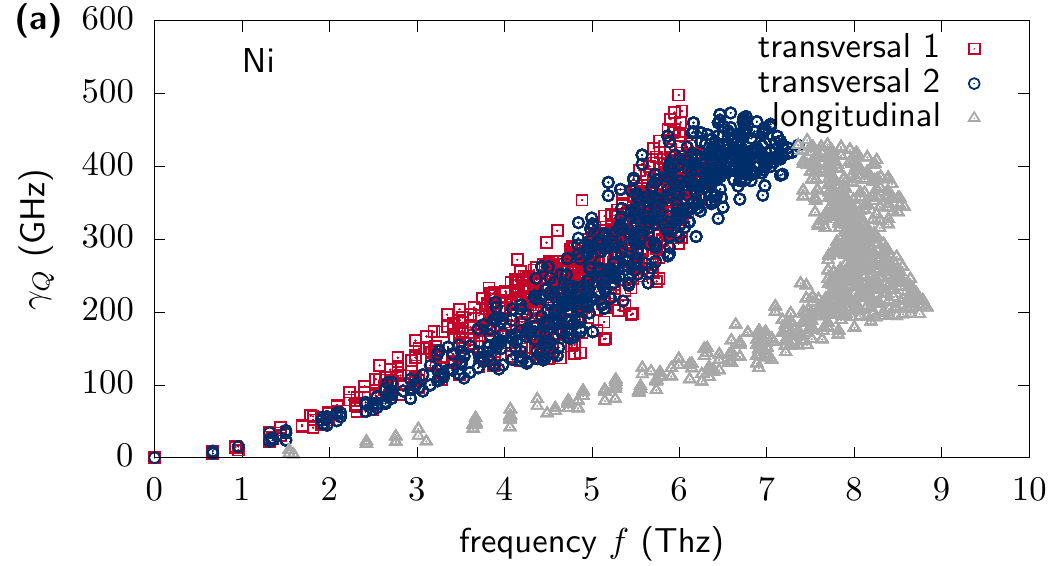}
	\includegraphics[width=0.48\textwidth]{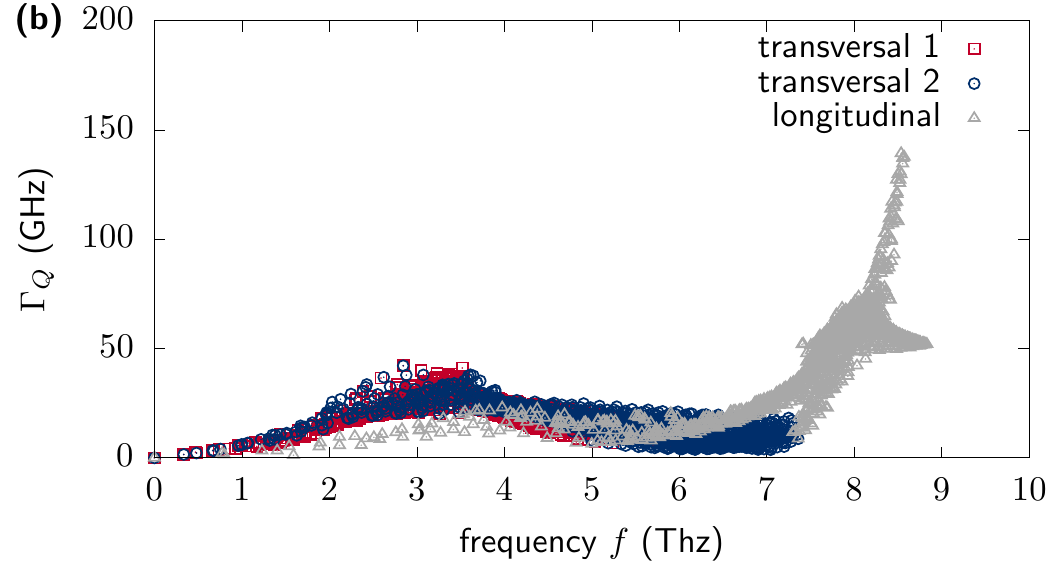}\\
	\includegraphics[width=0.48\textwidth]{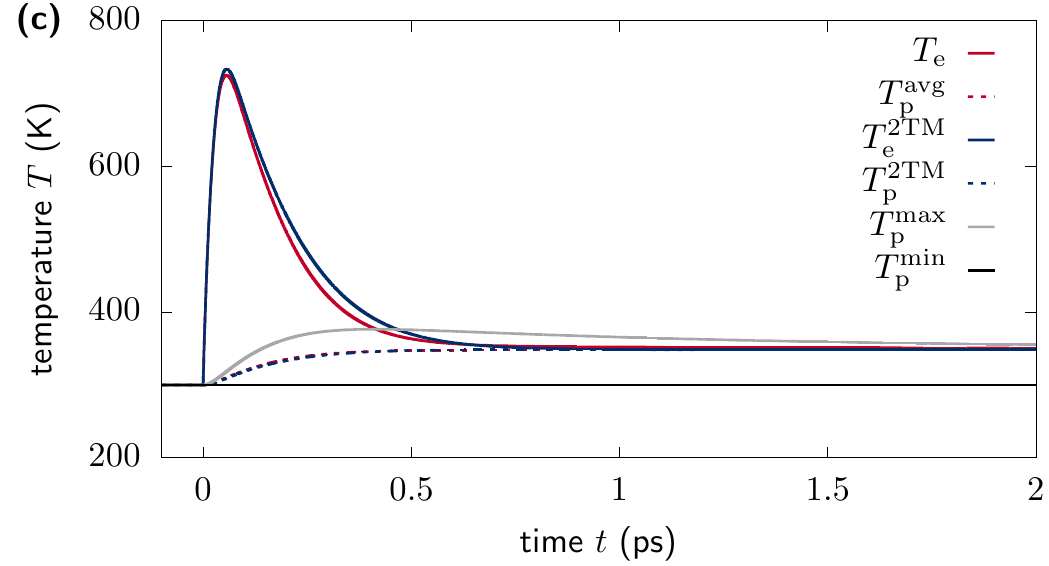}
	\caption{\label{fig_ni_2TM}(a) \textit{Ab initio} calculated electron-phonon linewidth dependent on the phonon frequency for the three phonon branches in nickel. (b) \textit{Ab initio} calculated phonon-phonon linewidth dependent on the phonon frequency for the three phonon branches in nickel. (c) Comparison of the computed temperatures  obtained from the two-temperature model with those of the \noneqmodel\ as a function of time after laser excitation.}
	\end{figure}%
		\begin{figure}[t]
	\centering
	\includegraphics[width=0.48\textwidth]{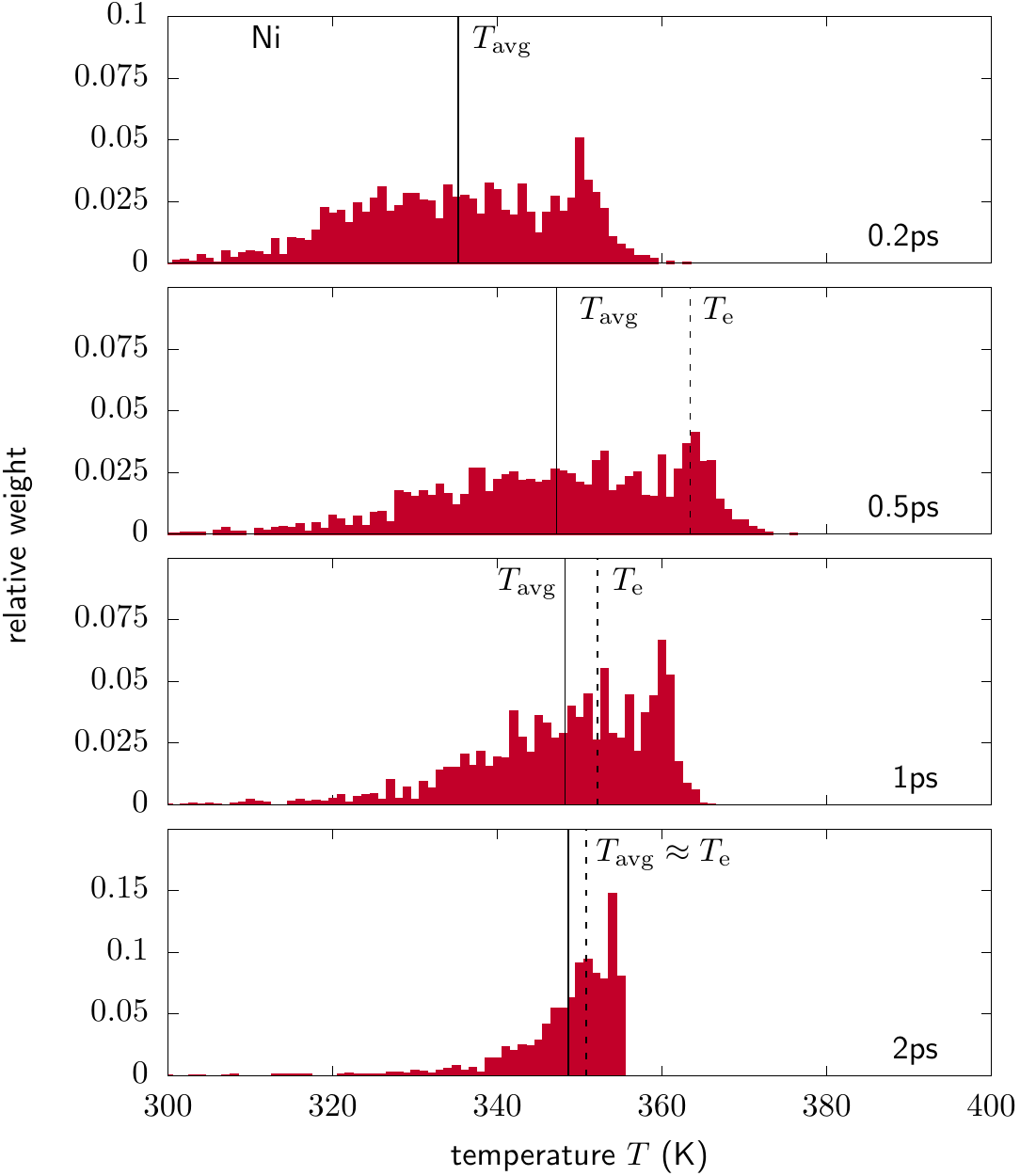}
	\caption{\label{fig_ni_hist}Phononic temperature distribution at different times during electron-phonon relaxation in nickel. The relative weight of phonon frequencies with a temperature within a range of 1K is shown during different times in the out-of-equilibrium states are shown.}
	\end{figure}%
The cases studied previously, for Au and Al, show the capabilities of our model to describe the out-of-equilibrium lattice dynamics of non-magnetic systems with very different electron-phonon and phonon-phonon couplings. In the following, we will extend the description to study the relaxation dynamics of ferromagnetic materials, specifically nickel, iron, and cobalt. First, we compute the phonon linewidths due to electron-phonon coupling  for the different materials by using spin-dependent densities of states and spin-dependent electron-phonon coupling. The mode-dependent linewidths in nickel are shown in Figure \ref{fig_ni_2TM}(a) and (b). The averaged linewidth due to electron-phonon coupling is 272.5\,GHz equivalent to a lifetime of about 1.8\,ps which is significantly larger than for gold and aluminium (see Table \ref{table_parameter}). The phonon-phonon linewidth in this material is on average only 27\,GHz leading to a lifetime of about 19 ps. 

In Figure \ref{fig_ni_2TM}(c), the temperature-evolution of electrons and phonons computed for the two-temperature model and the \noneqmodel\ are shown. Due to the strong electron-phonon coupling, the electrons are heated up within 100\,fs to about 750\,K and electrons and the averaged phonon temperature are nearly converged at about 0.75\,ps to a temperature of 349\,K. 
In Figure \ref{fig_ni_hist}, the effective phonon temperature distribution at different stages of the relaxation are shown. At 0.2\,ps, a broad distribution with a range of 60\,K is reached due to the mode-dependent coupling with the electrons. The maximum effective phonon temperature is much lower than in gold and aluminum and only about 380\,K. 
The maximum linewidth of electron-phonon coupling in aluminum is three times larger than the average value, whereas in nickel the maximum value is less than a factor 2 larger than the average. This leads to  narrower phonon  temperature distributions during the relaxation in nickel.

The thermalization of the phonons is mainly driven by electron-phonon coupling. Hotter phonons transfer their energy back to the electrons which are already colder than these phonons and this energy is again distributed to the less heated phonons. This is illustrated by the asymmetric temperature distributions at 1\,ps and 2\,ps. The peak of the phonon-temperature distribution is formed slightly above the electron temperature and it still deviates by several Kelvin from the averaged temperature. These results clearly demonstrate that the two-temperature model gives an  insufficient description of the temporal evolution of the lattice temperature, and it hides the physical mechanism that drives thermalization in Ni. Hence, although the results provided by the two-temperature model and the calculated averaged dynamics  of our model show a good agreement, the use of the two-temperature model leads to a wrong interpretation of the physical  phenomena involved in the lattice thermalization in Ni. 

\subsection{Out-of-equilibrium dynamics in iron}
\begin{figure}[t]
	\centering
	\includegraphics[width=0.48\textwidth]{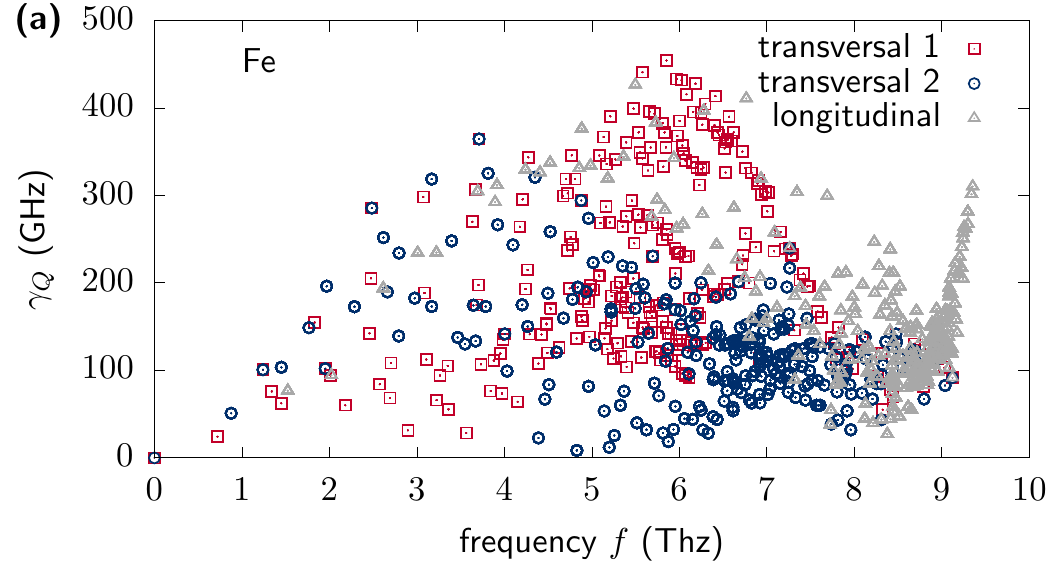}
	\includegraphics[width=0.48\textwidth]{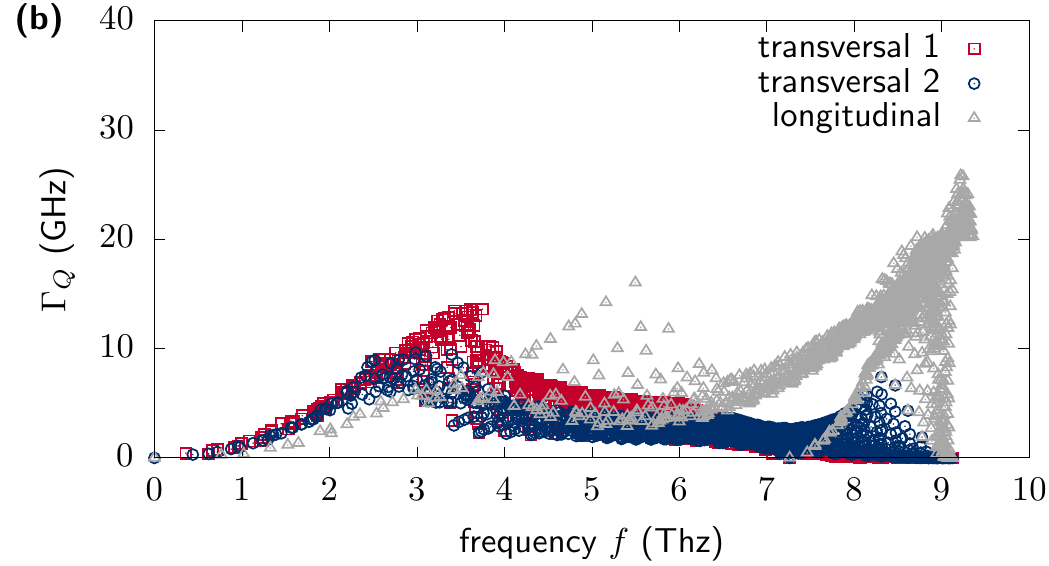}\\
	\includegraphics[width=0.48\textwidth]{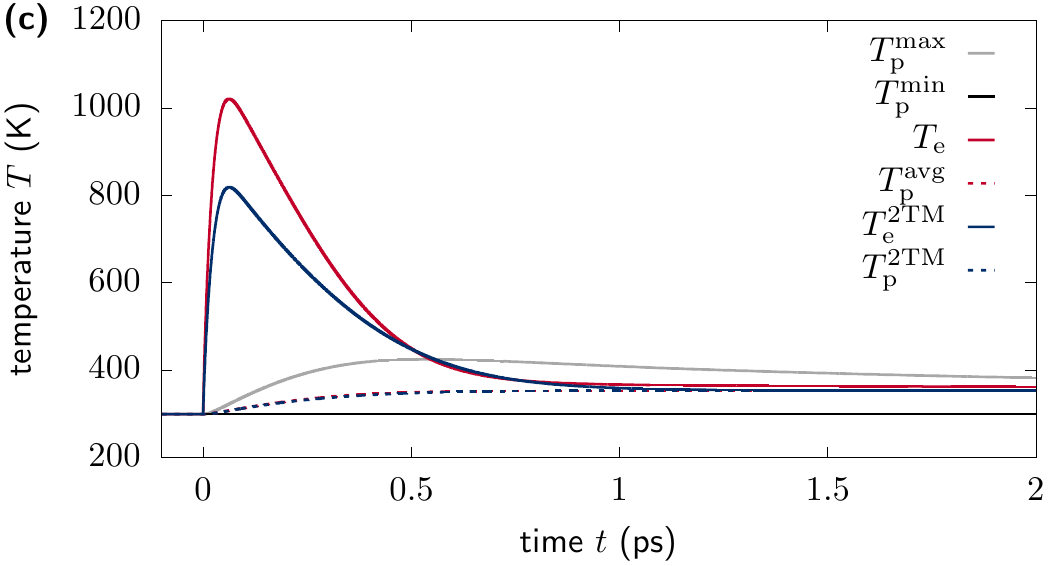}
	\caption{\label{fig_fe_2TM}(a) \textit{Ab initio} calculated electron-phonon linewidth dependent on the phonon  frequency for the three phonon branches  in iron. (b) \textit{Ab initio} calculated phonon-phonon linewidth dependent on the phonon  frequency for the three phonon branches in iron. (c) Comparison of the computed temperatures from the two-temperature model with  those from the \noneqmodel\ as a function of time after laser excitation.}
	\end{figure}	
	\begin{figure}[t]
	\centering
	\includegraphics[width=0.48\textwidth]{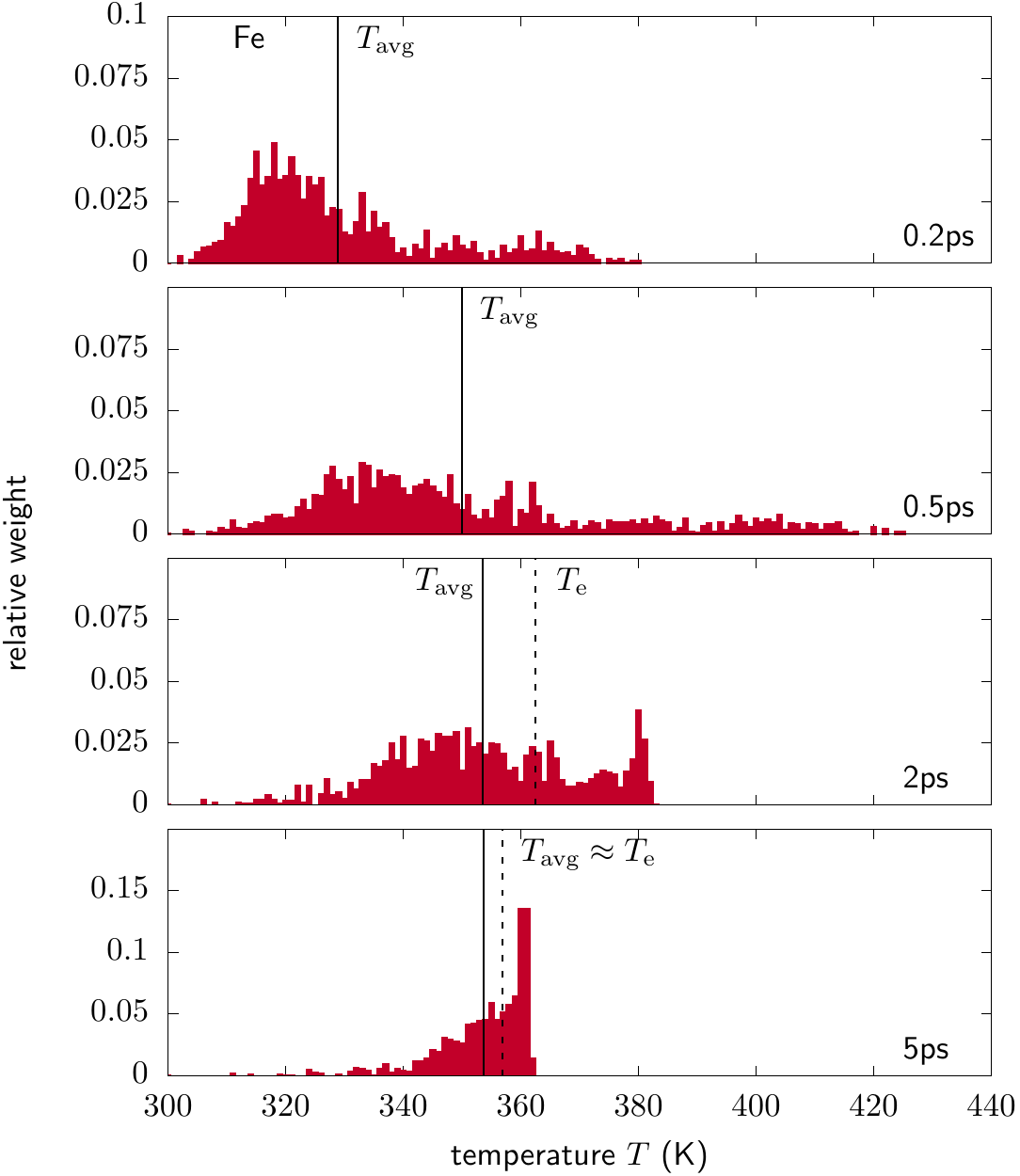}
	\caption{\label{fig_fe_hist}Phononic temperature distribution at different times during electron-phonon relaxation in iron. The relative weight of phonon frequencies  having  a temperature within an  interval of 1\,K is shown during different times in the out-of-equilibrium  dynamics.}
	\end{figure}
Iron presents the lowest phonon linewidths due to phonon-phonon interaction among the  here-discussed materials with an averaged value of 6.7 GHz (equivalent to a lifetime of 75 ps) as shown in Table \ref{table_parameter}. Additionally, the computed averaged electron-phonon coupling in iron is 164 GHz (equivalent to 3\,ps lifetime), suggesting a larger relevance of this scattering mechanism in the dynamics than the phonon-phonon interaction. More specifically, this is  illustrated in Figures \ref{fig_fe_2TM}(a) and (b), where the mode-dependent phonon linewidhts are shown due to electron-phonon and phonon-phonon scatterings, respectively.  The time evolution of the temperatures of electrons and phonons computed with the two-temperature model and the \noneqmodel\ are summarized in Figure \ref{fig_fe_2TM} (c). The results show that the  two-temperature model provides a dynamics that is in  significant
contrast with the out-of-equilibrium model, showing a faster relaxation with an electronic temperature reaching a maximum value of around 200 K lower than in our model. Despite this discrepancy it is noteworthy to observe that the averaged effective phonon temperature in our model shows a very good agreement with the lattice temperature from the two-temperature model.  


The distributions of the effective phonon temperatures during the relaxation are shown in Figure \ref{fig_fe_hist}. Unlike the case of nickel where the range of effective temperature deviations was about 70 K, in iron the range of temperatures is above 100 K due to a strongly phonon mode-dependent electron-phonon coupling. Moreover, the temperature distribution is very asymmetric with practically the same relative weight of phonons for all the temperatures spanned. As a consequence a large percentage of phonon modes have much larger temperatures than the averaged effective phonon temperature, and more importantly,  larger than the electron temperature. This results in a flow of energy from those phonon modes to the electronic system which then deliver the energy to the phonon modes with lower temperatures. Since the phonon-phonon coupling strength is small, it plays a small role in the lattice relaxation.  
Additionally, even though the averaged effective phonon temperature and the electron temperature are converged after 1\,ps, Figure \ref{fig_fe_hist} shows that the phonon system remains in an out-of-equilibrium state for longer times and only after more than 5\,ps, we can start considering that the system is in a quasi-equilibrium state, evidencing again the failure of the two-temperature model. 

\subsection{Out-of-equilibrium dynamics in cobalt}
\begin{figure}[t]
	\centering
	\includegraphics[width=0.48\textwidth]{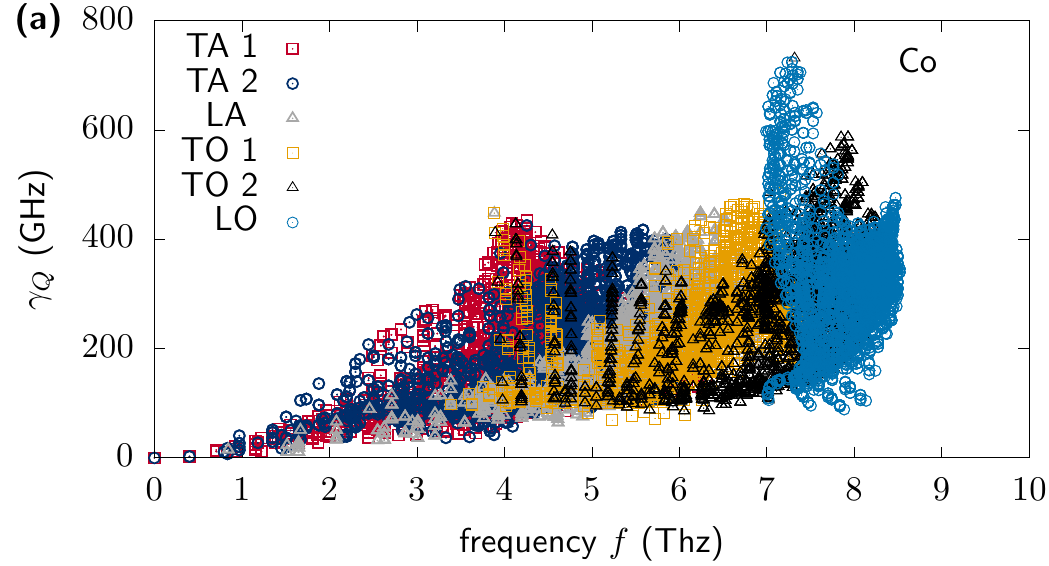}
	\includegraphics[width=0.48\textwidth]{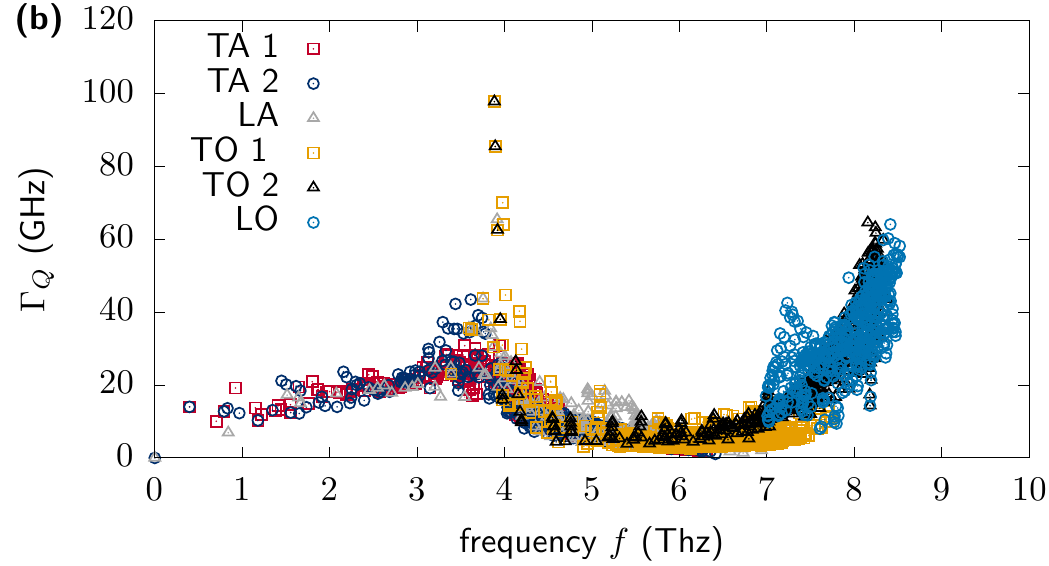}\\
	\includegraphics[width=0.48\textwidth]{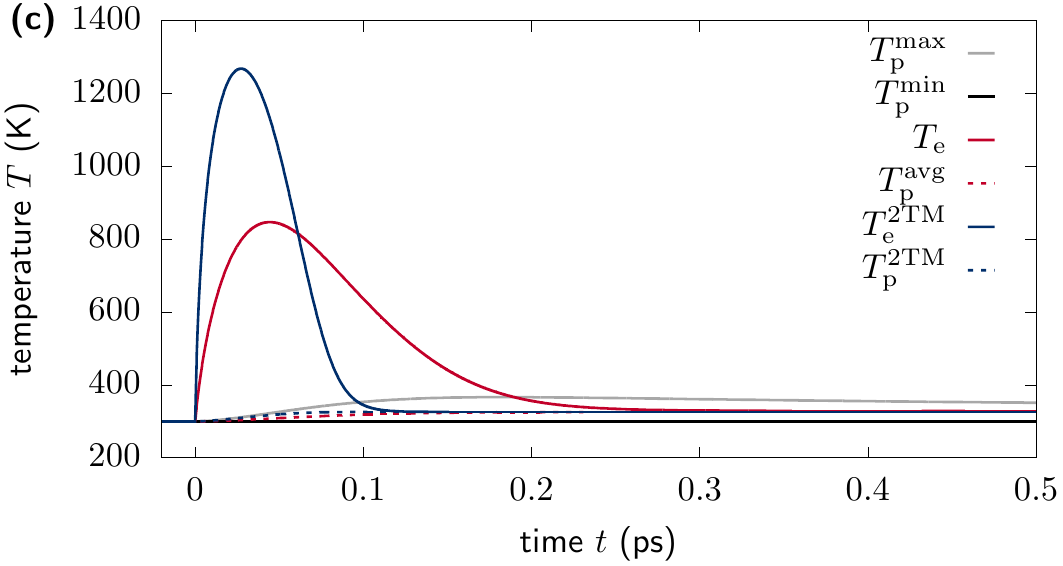}
	\caption{\label{fig_co_2TM}(a) \textit{Ab initio} calculated electron-phonon linewidth dependent on the phonon  frequency for the three phonon branches  in cobalt. (b) \textit{Ab initio} calculated phonon-phonon linewidth dependent on the phonon  frequency for the three phonon branches  in iron. (c) Comparison of the computed temperatures from the two-temperature model with  those from the \noneqmodel\ as a function of time after laser excitation.}
	\end{figure}	

As last example, we study the relaxation dynamics in hcp cobalt, which, as shown in Table \ref{table_parameter}, has the largest electron-phonon coupling strength among all the systems here analyzed. Moreover, hcp cobalt has two inequivalent atoms in the primitive cell, which leads to 6 different phonon branches instead of 3 as in the other materials. The calculated phonon linewidths due to  electron-phonon and phonon-phonon couplings are shown in Figure \ref{fig_co_2TM}(a) and (b). The averaged electron-phonon linewidth is about 244.4\,GHz corresponding to a lifetime of only 2\,ps and the averaged phonon-phonon linewidth is about 16.4\,GHz giving a lifetime of around 31\,ps. The larger complexity of the primitive unit cell in Co with respect to the other materials studied, favors a larger possibility for the electrons to interact with the phonons, leading to a very strong mode-dependent electron-phonon coupling.

The temporal evolution of the effective temperatures after laser excitation is shown in Figure \ref{fig_co_2TM}(c).  The two-temperature model deviates strongly from the \noneqmodel. The maximum electron temperature in the \noneqmodel\ remains below 900\,K, whereas the two-temperature model predicts a maximum electron temperature above 1200\,K. 
Significantly, the relaxation times of both models strongly deviate from each other.  The electron and phonon temperatures of the two-temperature model converge already shortly after 0.1\,ps, whereas the averaged effective phonon temperature and the electron temperature in the \noneqmodel\ converge at about 0.2\,ps.
The final temperature is much lower than in the other cases caused by the higher heat capacity of the phonons due to the larger number of phonon branches. 

\begin{figure}[!t]
	\centering
	\includegraphics[width=0.48\textwidth]{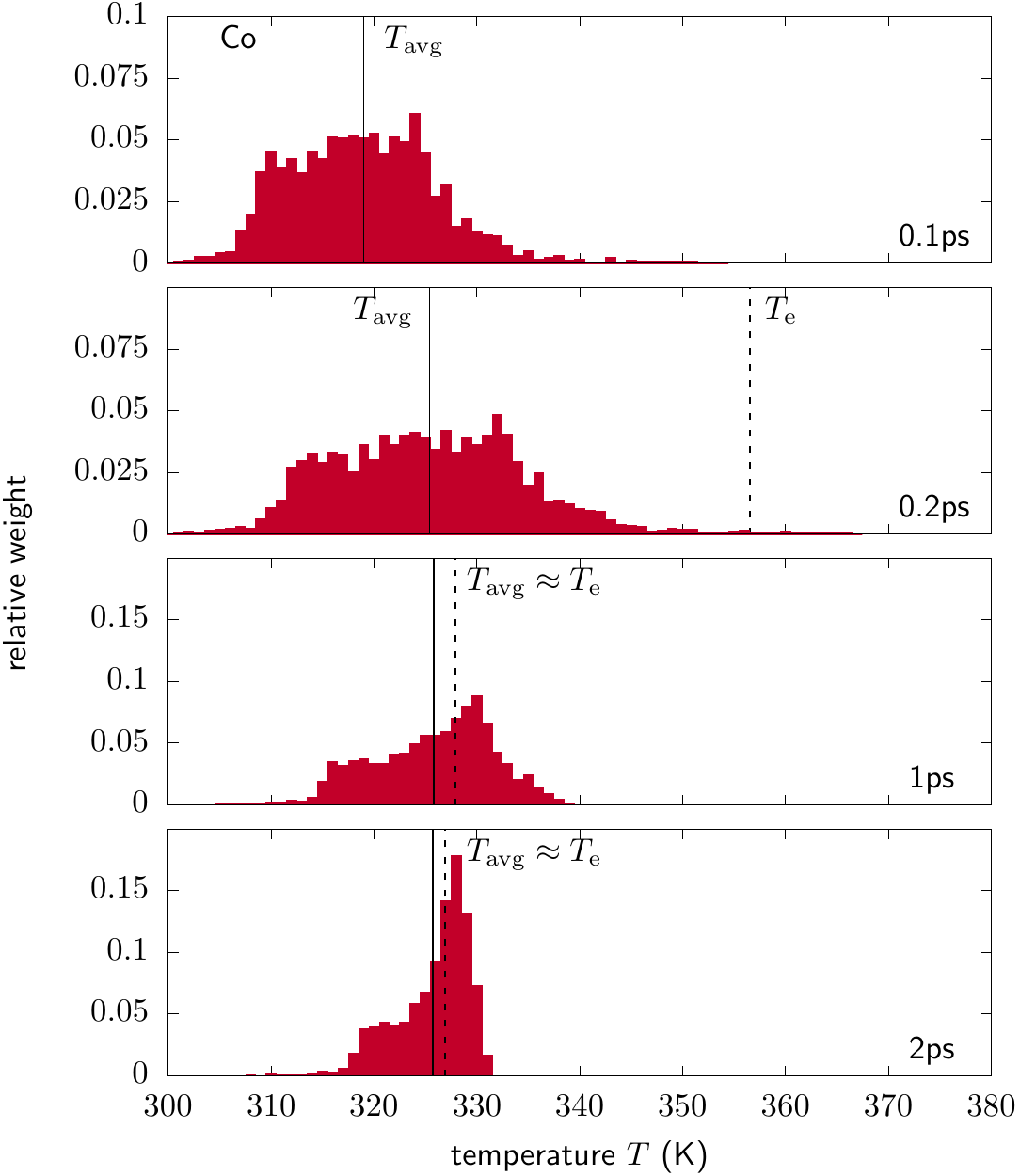}
	\caption{\label{fig_co_hist}Phononic temperature distribution at different times during electron-phonon relaxation in hpc cobalt. The relative weight of phonon frequencies having  a temperature within an interval of 1\,K is shown at different times.}
	\end{figure}

In contrast to the other two ferromagnetic metal systems, the effective phononic temperature distribution in cobalt is narrower as shown in Figure \ref{fig_co_hist}. The overall heating effect in the phononic system is smaller than for the other cases studied here due to the large phonon heat capacity. The broadest distribution is already obtained before 0.2\,ps and the effective  phonon temperature range is about almost 80\,K. It is important to mention that  during the timescale at which the system's evolution occurs the electrons remain out-of-equilibrium, and therefore the assumption of thermal electrons is not well justified and probably requires a more advanced description that is beyond the scope of this work. 

\section{Discussion and conclusions}
We have applied a  microscopic  out-of-equilibrium model to study the laser-induced  relaxation dynamics in five different metals, i.e., gold, aluminum, nickel, iron, and cobalt. As a general  observation, we have evidenced that the two-temperature model fails to describe  accurately the lattice dynamics and in some cases as in cobalt and iron, also the electronic dynamics. 
Surprisingly, this is evidenced even in the case of gold, where the assumptions on which  the two-temperature model  are based would seem better justified.

\begin{figure}[t]
\centering
\includegraphics[width=0.48\textwidth]{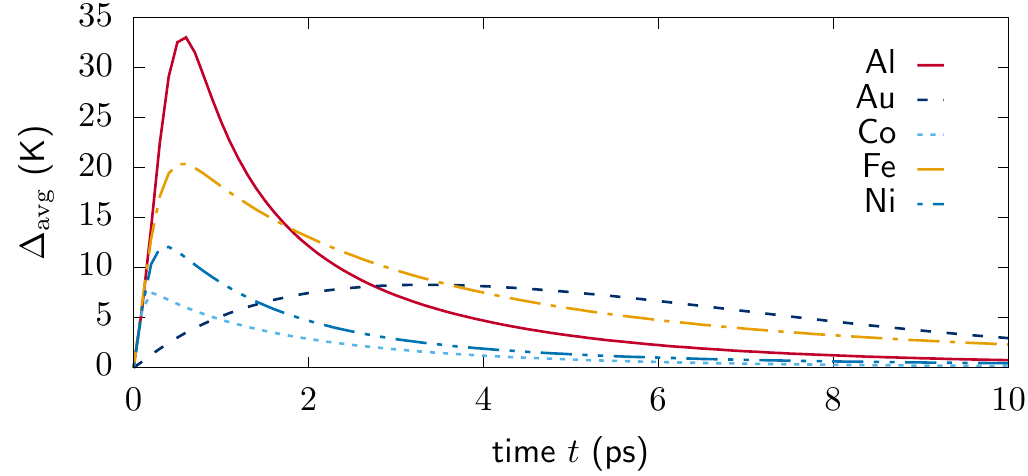}
\caption{\label{fig_noneq2}  Computed mean deviation from the time-dependent averaged  phonon temperature as a function of time for all materials.  
}
\end{figure}

To describe the out-of-equilibrium state more quantitatively we calculate the mean deviation from the time-dependent averaged effective phonon temperature $\Delta_\mathrm{avg}=\sum_Q(|T_Q-T_\mathrm{avg}|)/N_Q$. 
A comparison of the resulting deviation as a function of time is shown in Figure \ref{fig_noneq2}. In all cases the mean deviation first increases when the electron temperature is higher than the effective phonon temperatures and then decays on different timescales. Due to the strong phonon-phonon coupling in gold, the mean deviation does not exceed 10\,K, but due to the small electron-phonon coupling, deviations still appear on a long timescale larger than 10 picoseconds. The largest mean deviation is obtained in aluminum for which the mean deviation is around 35\,K after 1 ps of the laser excitation, evidencing a very spread distribution of phonon temperatures. 
In the ferromagnetic metals iron, nickel, and cobalt the mean deviation is always below 20\,K for the laser intensities used. In these cases, the averaged linewidth of electron-phonon coupling is larger than in aluminum, but the relative deviations from the averaged linewidth are smaller. Therefore, in aluminum we obtain larger values for the mean deviations indicating a state far from equilibrium, whereas in other materials the mean deviations are lower and the phononic systems are not driven as far from equilibrium. Nonetheless, this data underlines that the assumption of thermalized phonon distributions is not fulfilled on the picosecond timescale for all materials.

\begin{figure}[t]
\centering
\includegraphics[width=0.48\textwidth]{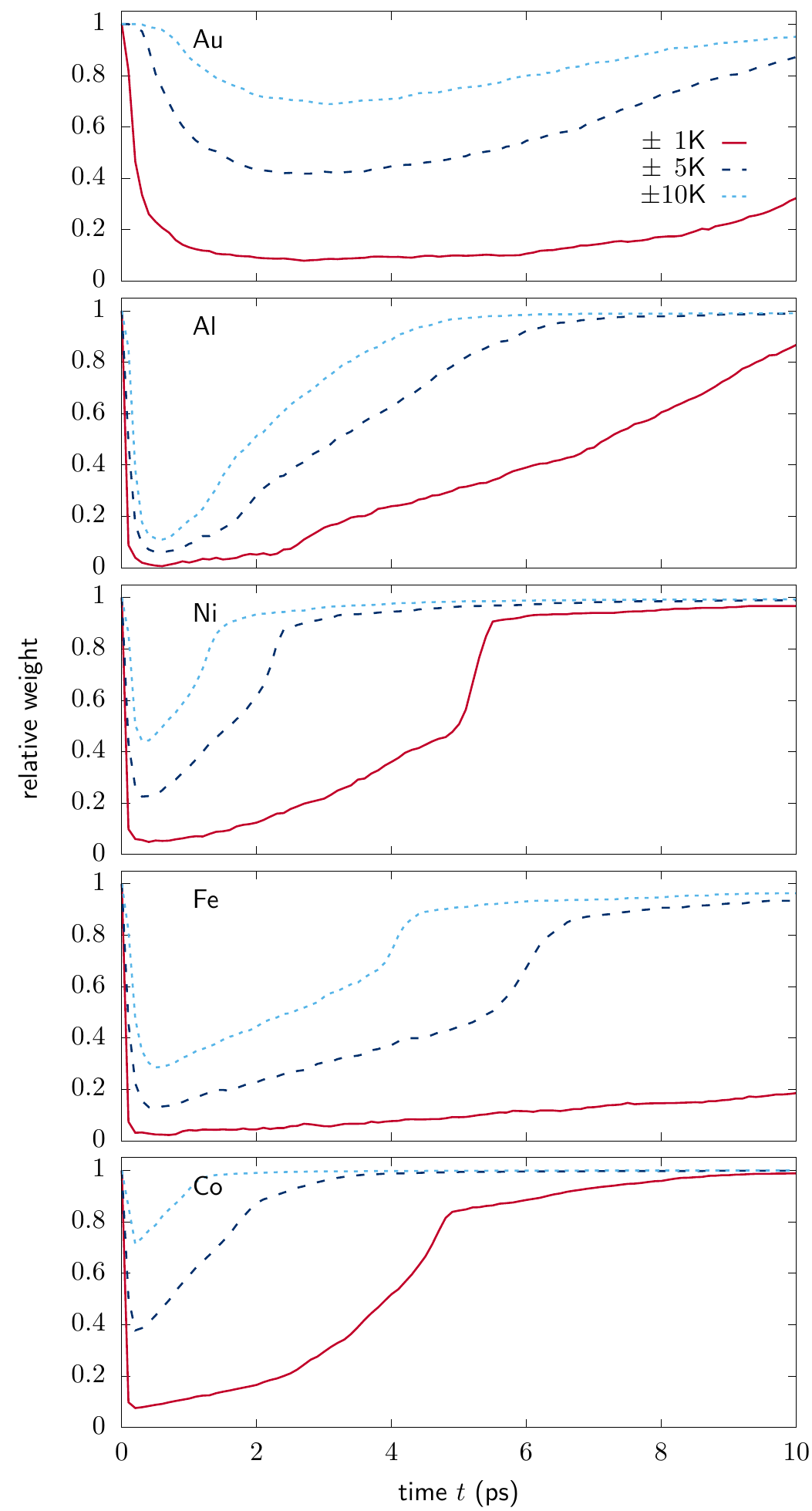}
\caption{\label{fig_noneq}Relative weight of phonon frequencies within a range of $\pm 1$\,K (red curve), $\pm 5$\,K (dark blue curve), $\pm 10$\,K (light blue curve) around the averaged temperature as a function of time for  the different metals considered.}
\end{figure}

As last step, we evaluate how good the description of the averaged  phonon temperature captures the overall behavior of the lattice relaxation. For that purpose we consider the relative amount of phonon modes around the averaged temperature within a range $\pm 1\,$K, $\pm 5\,$K, $\pm 10\,$K, which is shown for all materials in Figure \ref{fig_noneq}. The relative number of effective phonon temperatures around the averaged effective temperature first drops strongly and then smoothly recovers.
Aside the different timescales of the relaxation, the dynamical behavior itself differs between the materials.
The relative number of phonons within a range of 10\,K around the averaged temperature in gold drops to about 80\% which means that the averaged effective temperature represent the overall behavior of the majority of the modes. But in aluminum the value drops to only 15\% and the averaged temperature cannot be considered at all as good description for the system. Afterwards the value increases to almost 1 at around 5\,ps, whereas the number increases slower in gold and is still below 1 at 10 \,ps. 

In the ferromagnetic metals, we observe a different behavior. 
For iron and nickel the number of phonon modes around the averaged effective temperature with a difference up to 10\,K decreases below 50\% and recovers on different material specific timescales. 
This indicates that larger fractions of phonon modes have similar effective temperatures deviating from the averaged value. During thermalization the effective temperature of those modes converges towards the averaged temperature leading to a large increase of the relative amount of modes with effective temperature around the averaged value and indicates asymmetric distributions as shown for the ferromagnetic materials. In cobalt, the drop of the relative weight is less strong, since the total change of temperature is smaller due to the large heat capacity of the phonons.

In summary, we have shown strong out-of-equilibrium phononic distributions after laser excitation on the picosecond timescale demonstrating the limitations of the two-temperature model on that timescale. Our \noneqmodel\ based on \textit{ab initio} input parameter such as phonon mode dependent electron-phonon coupling and phonon-phonon coupling demonstrate that the obtained out-of-equilibrium dynamics are very material specific and the phonon mode dependences of the coupling processes have to be considered.  
In aluminum we have found the strongest deviations from equilibrium due to the decoupling of the different branches. It was proposed earlier by Waldecker \textit{et al}.\ to use a different coupling for each branch \cite{Waldecker2016}, which can improve the description of the dynamics. Our findings show that this works only in certain materials such as aluminum and although this can improve the description, it lacks still the strong mode-dependent electron-phonon coupling within each branch.

Our simulations clearly show that the temperature of the hottest phonons can exceed the electron temperature and relax on slower timescales than the electrons. This is in contradiction to a simplified model for electron-phonon relaxation by Sadasivam \textit{et al.}\ \cite{Sadasivam2017}, in which they assume that the hottest phonons would follow the electron temperature. This assumption excludes a possible energy redistribution of the highly excited phonon modes to lower excited ones via electron-phonon coupling. Through our explicit calculations, we demonstrate that this mechanism plays an important role in the case of strong electron-phonon coupling and weak phonon-phonon interaction.


All this  novel behavior demonstrates that the two-temperature model is an insufficient description of the relaxation process of electrons and phonons after laser excitation in the ps time range and mode-dependent dynamics have to be considered in order to describe the heating process and subsequent system dynamics on that timescale. 

\begin{acknowledgments}
We thank R. Ernstorfer for valuable discussions. The authors acknowledge financial support from the Deutsche Forschungsgesellschaft via RI 2891/1-1,  RI 2891/2-1, and  TRR 227 "Ultrafast Spin Dynamics", as well as from the Swedish Research Council (VR),  and the K.\ and A.\ Wallenberg Foundation  (Grant No.\ 2015.0060). We also acknowledge support from the Swedish National Infrastructure for Computing (SNIC).
\end{acknowledgments}

\appendix

\begin{widetext}
\section{Microscopic out-of-equilibrium dynamics model}

In the work by Maldonado \textit{et al.}\  \cite{Maldonado2017} the change of the phonon distribution function due to the phonon-phonon scattering was analytically derived, showing that it can be written as
%
\begin{align}
\dot{n}_Q\big|^{scatt.}_{ph-ph} =&\frac{2\pi}{\hbar}\sum\limits_{k,k'}|\Phi_{-Q,k,k'}|^2\Big[ \left(  n_Q(T_{\ell}^{kk'})-n_Q(T_{\ell}^Q)\right) (
n_k-n_{k'}) 
\left[\delta(\omega_Q+\omega_k-\omega_{k'})-\delta(\omega_Q-\omega_k+\omega_{k'})\right]   \nonumber\\
&  ~~~~~~ + \left(  n_Q(T_{\ell}^{kk'})-n_Q(T_{\ell}^Q)\right) \left(n_k+n_{k'}+1\right)\delta(\omega_Q-\omega_k-\omega_{k'}) \Big] \\
= &\frac{2\pi}{\hbar} \sum\limits_{k,k'}|\Phi_{-Q,k,k'}|^2 \left(  n_Q(T_{\ell}^{kk'})-n_Q(T_{\ell}^Q)\right) 
\Big[ (n_k-n_{k'})\left[ \delta(\omega_Q+\omega_k-\omega_{k'})-\delta(\omega_Q-\omega_k+\omega_{k'}) \right]  \nonumber\\
& ~~~~~~~ + (n_k+n_{k'}+1)\delta(\omega_Q-\omega_k-\omega_{k'})\Big] .
\end{align}
%
However, to reach this solution a single relaxation approximation was used, where it was assumed that $T_{\ell}^k\approx T_{\ell}^{k'}$. In this work, we extend the model by an exact description of  the phononic population rate. To do so, we start from the analytically exact expression derived by Maldonado \textit{et al.},
%
\begin{align}
\dot{n}_Q\big|^{scatt.}_{ph-ph}=&\frac{2\pi}{\hbar^2}\frac{36}{2}\sum\limits_{k,k'}|\Phi_{-Q,k,k'}|^2 \Big[(n_Q+1)(n_k+1)n_{k'}\delta(\omega_Q+\omega_{k'}-\omega_{k''})
+(n_Q+1)(n_{k'}+1)n_k\delta(\omega_Q+\omega_{k'}\nonumber\\ &-\omega_k)-n_Qn_k(n_{k'}+1)\delta(\omega_Q-\omega_{k'}+\omega_k)
-n_Qn_{k'}(n_k+1)\delta(\omega_Q+\omega_{k'}-\omega_k))\nonumber\\&+(n_Q+1)n_kn_{k'}\delta(\omega_Q-\omega_{k'}-\omega_k)
-n_Q(n_k+1)(n_{k'}+1)\delta(\omega_Q-\omega_{k'}-\omega_k)\Big] .
\label{phonvert}
\end{align}
%
%
The factor $\frac{36}{2}$ is introduced due to the 3! equivalent terms from the summation in the phonon-phonon matrix elements (square)  and the factor 1/2 to avoid double counting in the summation in equation \eqref{phonvert}. We can now use the following relations (and additionally, we define $\beta(k)\equiv \frac{\hbar\omega_k}{k_{\rm B}T_k}$)
\begin{align}
n_kn_{k'}=\frac{1}{(e^{\beta(k)}-1)(e^{\beta(k')}-1)}=(n_k+n_{k'}+1)\frac{1}{e^{\beta(k)+\beta(k')}-1}\label{nk1} ,\\
n_k(n_{k'}+1)=\frac{1}{e^{\beta(k)}-1}\left( \frac{1}{e^{\beta(k')}-1}+1 \right)=(n_k-n_{k'})\frac{-1}{e^{\beta(k)-\beta(k')}-1} , \label{nk2}
\end{align}
%
%
and then equations (\ref{nk1}) and (\ref{nk2}) become
\begin{align}
n_kn_{k'}=(n_k+n_{k'}+1)n_Q(\beta_0) , ~~~~~~~~~~\textrm{for} ~~~~ \delta(\omega_k-\omega_{k^{\prime}}-\omega_k)\\
n_k(n_{k'}+1)=-(n_k-n_{k'})n_Q(\beta_1),  ~~~~~~~~~~\textrm{for} ~~~~ \delta(\omega_k+\omega_{k^{\prime}}-\omega_k)\\ 
n_{k'}(n_{k}+1)=-(n_{k'}-n_{k})n_Q(\beta_2), ~~~~~~~~~~\textrm{for} ~~~~ \delta(\omega_k-\omega_{k^{\prime}}+\omega_k)
\end{align}
%
with
\begin{align}
\beta_0(k,k',Q)=\beta_0=\frac{\hbar \omega_k\left( T_{k'}-T_{k}\right)+\hbar\omega_QT_k}{k_{\rm B}T_kT_{k'}}=\frac{\hbar\omega_Q}{k_{\rm B}T_{k^{\prime}}}\left( 1+\frac{\omega_k}{\omega_Q}\frac{T_{k^{\prime}}-T_k}{T_k} \right)=\frac{\hbar\omega_Q}{k_{\rm B}T_{k}}\left( 1+\frac{\omega_{k^{\prime}}}{\omega_Q}\frac{T_k-T_{k^{\prime}}}{T_{k^{\prime}}}  \right) , \\
\beta_1(k,k',Q)=\beta_1=\frac{\hbar \omega_k\left( T_{k'}-T_{k}\right)+\hbar\omega_QT_k}{k_{\rm B}T_kT_{k'}}=\frac{\hbar\omega_Q}{k_{\rm B}T_{k^{\prime}}}\left( 1+\frac{\omega_k}{\omega_Q}\frac{T_{k^{\prime}}-T_k}{T_k} \right)=\frac{\hbar\omega_Q}{k_{\rm B}T_{k}}\left( 1+\frac{\omega_{k^{\prime}}}{\omega_Q}\frac{T_{k^{\prime}}-T_k}{T_{k^{\prime}}} \right) ,\\
\beta_2(k,k',Q)=\beta_2=\frac{\hbar \omega_k\left( T_{k}-T_{k'}\right)+\hbar\omega_QT_k}{k_{\rm B}T_kT_{k'}}=\frac{\hbar\omega_Q}{k_{\rm B}T_{k^{\prime}}}\left( 1-\frac{\omega_k}{\omega_Q}\frac{T_{k^{\prime}}-T_k}{T_k} \right)=\frac{\hbar\omega_Q}{k_{\rm B}T_{k}}\left( 1-\frac{\omega_{k^{\prime}}}{\omega_Q}\frac{T_{k^{\prime}}-T_k}{T_{k^{\prime}}} \right).
\end{align}
%
Finally, equation (\ref{phonvert}) can be written as
%
\begin{align}
\dot{n}_Q\big|^{scatt.}_{ph-ph} =& \,\frac{36\pi}{\hbar^2}\sum\limits_{k,k'}|\Phi_{-Q,k,k'}|^2\Big[  (n_k-n_{k'}) 
\left[\delta(\omega_Q+\omega_k-\omega_{k'})\left(  n(\beta_2)-n_Q\right)-\delta(\omega_Q-\omega_k+\omega_{k'}) \left(  n(\beta_1)-n_Q\right) \right]   \nonumber\\
&  ~~~~~~ + \left(  n(\beta_0)-n_Q\right) \left(n_k+n_{k'}+1\right)\delta(\omega_Q-\omega_k-\omega_{k'}) \Big]  \\
= &\, \frac{36\pi}{\hbar^2} \sum\limits_{k,k'}|\Phi_{-Q,k,k'}|^2  
\Big[ (n_k-n_{k'})\left( n(\beta_2)\delta(\omega_Q+\omega_k-\omega_{k'})-n(\beta_1)\delta(\omega_Q-\omega_k+\omega_{k'}) \right) \nonumber \\ &+n(\beta_0)(n_k+n_{k'}+1)\delta(\omega_Q-\omega_k-\omega_{k'})\Big] .
\end{align}
\end{widetext}

Then, depending on the cases considered, $\tilde{\Tp}^{k,k'}$ defined in equations \eqref{eq_dyn1} and \eqref{eq_dyn2} will be equal to $\tilde{\Tp}^{k,k'}= $$T_k\left( 1-\frac{\omega_{k^{\prime}}}{\omega_Q}\frac{T_{k^{\prime}}-T_k}{T_{k^{\prime}}} \right)^{-1}$ or $\tilde{\Tp}^{k,k'}=$ $T_k\left( 1+\frac{\omega_{k^{\prime}}}{\omega_Q}\frac{T_{k^{\prime}}-T_k}{T_{k^{\prime}}} \right)^{-1}$.

On the other hand, the phonon linewidth $\gamma_Q$ from electron-phonon coupling and phonon linewidth $\Gamma_{Qk'}$ from phonon-phonon scattering are given by (see also Ref.\ \cite{Maldonado2017}):
\begin{align}
    \label{eq_gamma}
	\gamma_Q=&\, 4\pi\omega_Q\sum_{k}{|M_{k,k'}|^2\delta(\epsilon_{k'}-\epsilon_\mathrm{F})\delta(\epsilon_k-\epsilon_\mathrm{F})} ,\\
	\label{eq_gamma_ph}
	\Gamma_{Q,k}=&\, \frac{18\pi}{\hbar^2}|\Phi_{-Q,k,k'}|^2 \big[ (n_k-n_{k'})\nonumber\\&\times (\delta(\omega_Q+\omega_k-\omega_{k'})-\delta(\omega_Q-\omega_k+\omega_{k'}))\nonumber\\&+(n_k+n_{k'}+1)\delta(\omega_Q-\omega_k-\omega_{k'}) \big].
	\end{align}
Here $|M_{k,k'}|$ and  $|\Phi_{-Q,k,k'}|$ denote electron-phonon and phonon-phonon matrix elements and $\epsilon_{k'}$ is the energy of the electron with momentum $k'$ and $\epsilon_\mathrm{F}$ the Fermi energy.

Quantity $I(\Te)$  introduced in Eqs.\ (\ref{eq_dyn1}) and (\ref{eq_dyn2}) is a correction factor for the electron-phonon coupling accounting for scattering of electrons away from the Fermi surface. It can be computed by: 
	\begin{align}
		I(\Te)=-\int_{-\infty}^{\infty}{\mathrm{d}\epsilon\partialdif{f_k}{\epsilon}\frac{g(\epsilon)^2}{g(\epsilon_\mathrm{F})^2}}\mathrm{,}
	\end{align}
where $g(\epsilon)$ is the electron density of states at energy $\epsilon$.

The quantity  $J(\omega_Q, \Tp^Q)$  in Eqs.\ (\ref{eq_dyn1}) and (\ref{eq_dyn2}) represents the second-order term in a Taylor expansion in temperature differences of the out of equilibrium phonon population
	\begin{align}
	J(\omega_Q, \Tp^Q)=\frac{\hbar\omega_Q}{k_\mathrm{B}\Tp^2}\left(\frac{\exp{\left(\frac{\hbar\omega_Q}{k_\mathrm{B}\Tp^Q}\right)}+1}{\exp{\left(\frac{\hbar\omega_Q}{k_\mathrm{B}\Tp^Q}\right)}-1}-\frac{2k_\mathrm{B}\Tp^Q}{\hbar\omega_Q}\right).
	\end{align}

\vspace*{1.5cm}
\bibliography{./library}

\end{document}